# Reading Users' Minds from What They Say:
# An Investigation into LLM-based Empathic Mental Inference

A Preprint

**Qihao Zhu**
Singapore University of Technology and Design

**Leah Chong**
Massachusetts Institute of Technology

**Maria Yang**
Massachusetts Institute of Technology

**Jianxi Luo**
City University of Hong Kong

March 20, 2024

## Abstract

In human-centered design, developing a comprehensive and in-depth understanding of user experiences—empathic understanding—is paramount for designing products that truly meet human needs. Nevertheless, accurately comprehending the real underlying mental states of a large human population remains a significant challenge today. This difficulty mainly arises from the trade-off between depth and scale of user experience research: gaining in-depth insights from a small group of users does not easily scale to a larger population, and vice versa. This paper investigates the use of Large Language Models (LLMs) for performing mental inference tasks, specifically inferring users' underlying goals and fundamental psychological needs (FPNs). Baseline and benchmark datasets were collected from human users and designers to develop an empathic accuracy metric for measuring the mental inference performance of LLMs. The empathic accuracy of inferring goals and FPNs of different LLMs with varied zero-shot prompt engineering techniques are experimented against that of human designers. Experimental results suggest that LLMs can infer and understand the underlying goals and FPNs of users with performance comparable to that of human designers, suggesting a promising avenue for enhancing the scalability of empathic design approaches through the integration of advanced artificial intelligence technologies. This work has the potential to significantly augment the toolkit available to designers during human-centered design, enabling the development of both large-scale and in-depth understanding of users' experiences.

## 1 Background and Introduction

Developing a comprehensive and deep understanding of people's experiences, i.e., empathic understanding [1], is crucial in human-centered design for accurately identifying and designing for human needs [2]. Developing such understanding typically involves engaging with users through methods like interviews [3] and in-depth observation of contextual inquiries [4]. However, these approaches, typically focusing on small groups of users, may not fully capture the views of the broader population. Additionally, the ability of designers to develop empathic understanding varies among individuals and depends on the extent in which they share the same experiences and cultural background as the users [5, 6]. Conversely, data-driven approaches, which gather and analyze large amounts of user-generated content (UGC) using natural language processing (NLP) techniques, have been extensively explored [7-9]. While these approaches are independent of designers' empathic ability and ensure reflections of the majority's opinions and preferences, they struggle to attain an in-depth understanding of users' underlying mental contents. Such advantages and disadvantages of traditional versus data-driven methods highlight the trade-off between scale and depth in user research [10].

The recent advancements in artificial intelligence (AI) have prompted discussions on artificial empathy (AE), aiming to equip AI with human-like empathic capabilities. Although AE has been explored in the AI and robotics fields for many years [11-13], its introduction to the field of design is relatively recent. Zhu and Luo [14] have proposed a high-level methodological framework of AE for human-centered design, combining theory-driven and data-driven perspectives on empathy modeling. This framework identifies key modules and components for breaking down the complex processes of empathic understanding in design and emphasizes the development and validation of AE systems at the component level.

Mental inference is a crucial component of AE that involves complex cognitive mechanisms to understand the tacit and implicit knowledge of people's mental contents (e.g., emotions, intentions, motivations, goals, and needs). This ability is vital for an artificial system to transcend mere analysis of explicit expressions and develop a deeper empathic understanding of people. Therefore, when AE can successfully perform mental inference, it holds the potential to enhance the user research process.

Furthermore, the design field lacks standard metrics and baseline datasets for evaluating the mental inference capabilities of AE systems in human-centered design. Recent research has investigated designers' performance in inferring users' mental states often through empathic accuracy tests [1, 15, 16]. These tests measure the similarity between designers' inferred mental states and users' self-reported ones, greater similarity indicating better empathic understanding. There is potential to adapt these tests as automatic metrics for AE's inference performance by collecting baseline data of users' self-reported mental states and benchmarks of designers' inferences [14].

Considering this background, the research question we aim to investigate in this paper is: "Can AE infer the underlying mental contents of users with performance comparable to that of human designers?" We (1) propose using Large Language Models (LLMs) to infer underlying mental contents from user comments and (2) develop empathic accuracy metrics for AE's performance in mental inference and use them to evaluate the proposed method. The codes for LLM inference and evaluation, as well as the involved data, are available on GitHub: https://github.com/Qihao-Zhu/LLM-Mental-Inference-Experiment

## 2 Methods

To address the research question, we design and conduct an experiment aiming at measuring and comparing the abilities of human designers and LLMs to infer the motivation-driven mental states of users. This section outlines the theoretical foundation of the mental inference tasks in the experiment, describes our LLM-based approaches for mental inference, and details the evaluation metric adapted from empathic design research.

### 2.1 Mental Inference Tasks

Mental contents, describing the underlying mental states of individuals, encompass a wide range of constructs including goals, needs, beliefs, desires, motivations, intentions, and more. Throughout the history of cognitive psychology, many theories have been developed around these concepts and their interrelations [17-20]. Activity Theory posited that human activities are motivated by the pursuit of socially and culturally constructed goals, which can be organized hierarchically [17, 21]. Building upon this theory, Hassenzahl [22] introduced a goal hierarchy to elucidate users' experiences in goal-directed interactions with products, comprising three levels: do-goals, motor-goals, and be-goals. The do-goal is the intermediate level in the goal hierarchy which refers to the specific tasks the users aim to achieve or complete. The motor-goal is product-specific and refers to the concrete sub-actions or steps they need to take regarding the use of a product to achieve a do-goal. On the other hand, the be-goal is more self-referential and refers to the users' overarching goals that are related to their broader life aspirations, self-perception, and/or desires about their own identities. Additionally, Self-Determination Theory proposes that human motivation is fueled by innate psychological needs [20]. Psychological needs, according to Deci and Ryan [20], are "Human needs specify innate psychological nutriments that are essential for ongoing psychological growth, integrity, and well-being". Psychological needs are considered the most self-referential components of our motivational mental states that drive our interaction with a product. From this theory, Desmet and Fokkinga [23] identified 13 fundamental psychological needs (FPNs) as essential for user experience, informing design research and practice. These FPNs include autonomy, beauty, comfort, community, competence, fitness, impact, morality, purpose, recognition, relatedness, security, and stimulation.

In our study, we focus on two constructs of mental contents: goals and psychological needs, as they are directly involved in the motivational system behind humans' product use. We adopt Hassenzahl's [22] three-level hierarchy of goals to categorize users' underlying goals related to a product, and Desmet and Fokkinga's [23] typology of 13 FPNs to identify foundational and self-referential factors motivating user experiences. Considering that Self-Determination Theory indicated the pursuit of goals is motivated by psychological needs [20], we integrate the three levels of goals with the FPNs into a comprehensive spectrum of mental states, ranging from product-specific to self-referential, thereby representing a full spectrum of empathic understanding of users' motivation-driven experiences. This spectrum is depicted in Figure 1, adapted from Hassenzahl [22].

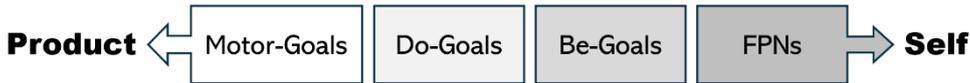

**Figure 1. Spectrum of motivation-directed mental contents, adapted from Hassenzahl [22]**

The experiment in our paper incorporates two mental inference tasks: goals inference and FPNs attribution. In the goals inference task, human subjects analyze user comments to discern underlying product-related goals: do-goals, motor-goals, and be-goals as outlined by Hassenzahl [22]. The FPNs attribution task requires the subjects to attribute the inferred goals to the 13 FPNs using a 5-point Likert scale, where 1 indicates "not attributed at all" and 5 denotes "highly attributed." These two tasks are also performed by LLM-based AI models, and their performance is compared to human subjects' performance.

**2.2 LLM-based Mental Inference**

The recent advancements in LLMs have shown promising capabilities in theory-of-mind (ToM) [24, 25], a fundamental cognitive mechanism underlying human empathy that involves inferring and understanding others' mental states [26]. While these developments indicate LLMs' potential in ToM, their evaluations have predominantly involved the false-belief test [24, 25]. This task, derived from cognitive psychology, tests artificial agents' understanding of fictional characters' false beliefs regarding their situations. Although this test is pivotal for assessing ToM in humans and LLMs, its fictional nature may not yield directly applicable insights for the context-dependent nature of design research and practice [15].

In our work, we leverage LLMs to perform mental inference tasks, aiming to understand users' motor-goals, do-goals, be-goals, and FPNs from their product-related comments. We explored the application of LLMs through three different zero-shot prompt engineering approaches: standard, chain-of-thoughts (CoT), and tree-of-thoughts (ToT). The standard approach employs direct instruction techniques commonly used in conversational LLMs, such as ChatGPT, where LLMs are queried with three specific questions derived from Hassenzahl [22] regarding the different types of goals.

Separately, the LLMs are tasked with mapping the user's comments to the 13 FPNs on a scale from 1 ("not attributed at all") to 5 ("highly attributed").

The CoT prompting approach [27], designed to break down complex tasks into simpler, intermediate steps for guided reasoning, has shown efficacy in enhancing LLMs' performance on cognitively demanding problems [28]. For the CoT setting in our experiment, an added step requires the LLM to adopt the user's perspective and envision the full context of their experience with the product. The LLM's reasoning process is then linearly structured from perspective-taking to goal inference and subsequently to FPNs attribution, incorporating the responses from preceding steps into subsequent queries.

Expanding on CoT, the ToT approach [29] introduces multiple reasoning paths by employing a high temperature setting, encouraging the generation of varied responses at each step. This method mirrors the human process of navigating through a combinatorial space of thoughts. An additional voting mechanism is utilized to determine the most appropriate response, which is then used to inform the subsequent step of inference. The voting function for perspective-taking and goal inference tasks are directly adapted from Yao et al. [29]. For FPNs attribution, rather than selecting an overall top response, the method aggregates commonalities across attempts, assigning the most consistent rating to each FPN based on agreement. For instance, if among five attempts to attribute "Autonomy," three suggest a rating of 3, then this rating is adopted as the consensus for "Autonomy." The implementation of these three prompting strategies in our study is visualized in Figure 2. The detailed prompt engineering workflow along with the prompt templates used can be found in our GitHub repository.

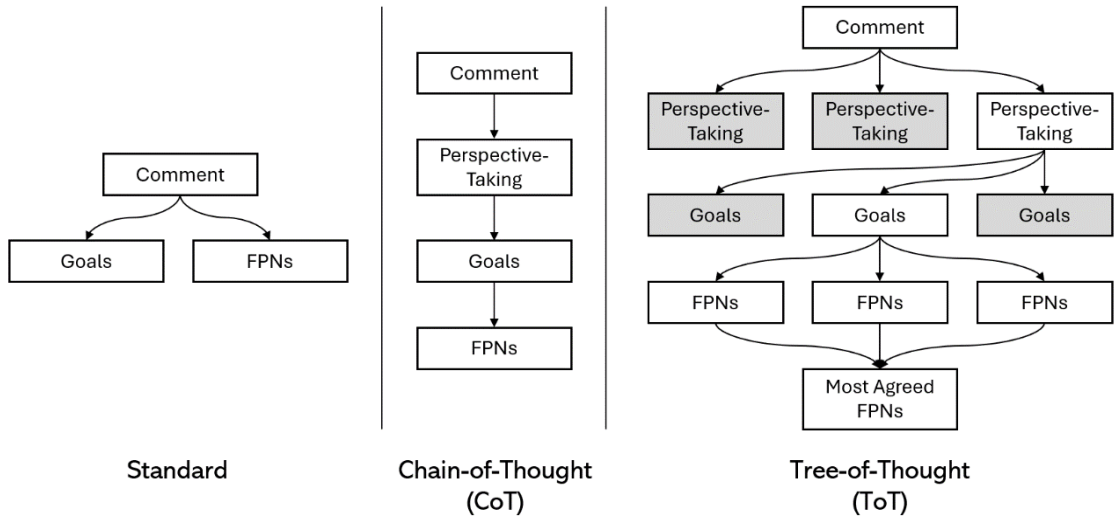

**Figure 2. Schematic illustration of prompt engineering approaches used in this paper, adapted from Yao et al. [29]**

### 2.3 Empathic Accuracy Metric

The design research community has explored various methods for measuring designers' empathic performance. Surma-Aho & Hölttä-Otto [1] identified six categories of empathy measurements: empathic tendency, beliefs about empathy, emotion recognition, understanding mental contents, shared feeling, and prosocial responding. Among these, "understanding mental contents" is deemed the most direct measure for evaluating the empathic understanding developed by designers. The empathic accuracy test, borrowed from cognitive psychology, is a proven method in empathic design for assessing the accuracy of understanding users' mental contents [15]. It compares the mental contents inferred by designers with those self-reported by users, with higher similarity scores indicating a more accurate understanding of the user.

As outlined in Section 1, the goal of our study is to evaluate the performance of AE in inferring mental contents, comparing it against human designers tasked with the same set of questions. For this purpose, we adopt the empathic accuracy test and develop an empathic accuracy metric for AE, requiring both users' self-reported mental contents for a baseline and designers' inferences as a benchmark [14]. Figure 3 illustrates the experimental schemes for developing the empathic accuracy metric and evaluating the performance of LLMs based on this metric.

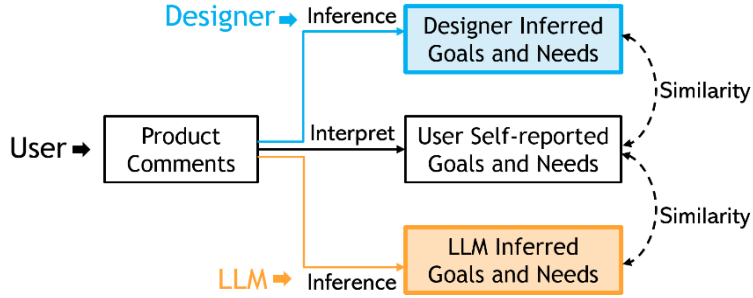

**Figure 3. Experiment schemes**

The data collection process unfolds in two stages, each stage involving user and designer participants respectively.

In the first stage, the user participants are asked to think about a recently purchased product or one they intend to buy soon. They then write a comment detailing their experience with the product (if already purchased) or their expectations of the product (if intended for future purchase), with no restrictions on comment length. Subsequently, the user participants are asked to familiarize themselves with the 13 FPNs and their sub-needs via a brochure developed by Desmet and Fokkinga [30]. Once they are familiar, they are tasked to identify their underlying product-related goals and map these goals to the appropriate FPNs based on their experiences or expectations.

In the second stage, the comments collected from the user participants are presented to the designer group. Each designer participant receives 6 comments. The designers, having learned the same FPNs material as the user participants, are instructed to infer the users' underlying goals and attribute them to the FPNs.

The accuracy of inferences made by human designers or LLMs against users' self-reported goals and needs is quantified using cosine similarity. Cosine similarity measures the angle between two vectors, resulting in a similarity score ranging from -1 to 1, where -1 indicates the two vectors are diametrically opposite and 1 indicates they are identical. It is frequently used in AI research to measure the semantic similarity between vector embeddings extracted from texts. To ensure the semantic meanings are accurately presented, the texts describing the goals are proofread first to correct any grammatical or spelling errors. These corrected texts are then transformed into vector representations using a text embedding model, allowing the computation of cosine similarity to gauge semantic similarity. For FPN attributions, the data are reported in a 5-point Likert scale, resulting in a 13-dimensional vector for each inference, which is also compared using cosine similarity. This approach ensures that the evaluation of AE's performance in inferring users' mental contents is grounded in empirical data, comparing it directly to the performance of human designers.

## 3 Data Collection

n the initial stage of our study, we recruited 27 participants to provide user feedback. The participants were recruited through mailing lists of non-design-focused departments in Massachusetts Institute of Technology or flyers that we posted around campus. Each participant was involved in a 30-minute session, which could be conducted either in-person or virtually via Zoom by their own preference.

They were compensated with a free meal (worth about $13) for their participation. They were instructed to provide one comment about a product and interpret their goals and FPNs related to that product following instructions. To ensure a diverse range of product types, participants were allocated into one of three broad activity categories: house cleaning (vacuum, broom, window cleaner, etc.), traveling (luggage, travel adapter, portable devices, etc.), and digital entertainment (headphones, smartphone, gaming console, etc.). After a preliminary review, we discarded one response due to its poor quality and selected the first 8 responses from each activity category, culminating in a dataset of 24 user comments, along with the baseline goals and FPN interpretations generated by the users.

In the subsequent stage, we enlisted 25 designers for in-person study sessions lasting an hour each. The participants are recruited through mailing lists of design-focused departments (e.g., mechanical engineering, architecture) at Massachusetts Institute of Technology or flyers that we posted around campus. We specifically required that the participants should have design-related backgrounds when they signed up. Each participant was compensated with $20. The study sessions required participants to review 6 different user comments and infer the respective users' goals and FPNs. Upon reviewing the collected data, we excluded the responses from one designer who failed to adhere to our instructions. From the remaining contributions, we compiled the first 5 inferences for each comment into a benchmark dataset, which includes responses from 20 designers, amounting to 120 data points in total. The study has been exempted from the Institutional Review Board (IRB) at Massachusetts Institute of Technology. All participants, including users and designers, signed an informed consent before participating. Table 1 presents the demographic information of participants from both groups whose responses were included in the final datasets.

Table 1. Information about participants

| Demographics of both user and designer participants | | | |
|---|---|---|---|
| | | User participants (24 in total) | Designer participants (20 in total) |
| Age | 18~25 | 66.7% | 70.0% |
| | 26~41 | 29.2% | 30.0% |
| | 42~57 | 4.2% | 0.0% |
| Gender | Female | 75.0% | 65.0% |
| | Male | 25.0% | 35.0% |
| Education level | Below college | 4.2% | 10.0% |
| | College but no degree | 33.3% | 15.0% |
| | Bachelor's degree | 45.8% | 35.0% |
| | Graduate or professional degree | 16.7% | 40.0% |
| Expertise experience of designer participants | | | |
| | | Design Experience | User research experience |
| Designer experience level | No Experience | 0.0% | 20.0% |
| | Little experience | 5.0% | 25.0% |
| | Some experience | 60.0% | 45.0% |
| | A lot of experience | 30.0% | 5.0% |
| | Expert | 5.0% | 5.0% |

Simultaneously, we gathered inference responses from Large Language Models (LLMs). Employing the three prompt engineering techniques outlined in Section 2.2, we conducted experiments with two state-of-the-art LLMs from OpenAI: GPT-3.5-turbo and GPT-4, creating a total of 6 LLM experimental groups. At the time of our study, GPT-4 was the most advanced model available with the largest model size and complex architecture [31]. The specific model versions used were "gpt-3.5-turbo-0613" and "gpt-4-0613," accessed via the OpenAI API using Python. For the standard and CoT prompting experimental groups, we set the temperature parameter to 0 to fixate the model's responses for consistent reproducibility. However, the ToT approach necessitated a higher temperature setting to facilitate the generation of multiple choices of response. In the ToT experimental groups, we used a temperature of 0.7 to generate 5 choices for each step and conducted 5 times of voting to determine the best choice. Each user comment was processed by each LLM group once, resulting in a total of 144 LLM inference data points.

## 4 Results

Following the collection of user baseline, designer inference, and LLM inference data, the analysis was conducted by: (1) calculating similarity scores between the designers' inferences and the user baseline to benchmark human designer performance in understanding users' minds; (2) calculating similarity scores between LLMs' inferences and the user baseline to assess LLM performance in mental inference; and (3) comparing LLM performance with the human designer benchmark based on similarity scores to the user baseline. For textual response embeddings, we utilized OpenAI's "text-embedding-3-large" model, recognized for its high performance on the Massive Text Embedding Benchmark [32]. This model generates a 3072-dimensional vector representation for each goal response, capturing its semantic meaning. For FPNs attributions, each answered in a 1 to 5 scale, we normalized these scores to a 0 to 1 scale before computing cosine similarity. Table 2 presents a user comment example alongside examples of designer and LLM inferred goals based on this comment. The float number below each goal type indicates its cosine similarity score to the baseline. In the example, both human designers and the LLM achieved relatively high similarity scores in inferring users' underlying goals.

Figure 4 details the overall performance of human designers and various LLM experimental groups in inferring different types of mental contents. Three key observations emerge from the analysis: (1) the performance of GPT-4 model matches human designers across all prompting techniques and is even found surpassing human designers in some of the goal inference tasks; (2) the GPT-3.5 model performs well in goals inference tasks but faces challenges when inferring attributing FPNs, with some outlier instances of non-compliance with instructions noted. These findings suggest that LLMs can reach human-level performance in mental inference. One interesting observation to take note is that, CoT and ToT prompting methods do not seem to enhance the overall performance of LLM in the mental inference tasks.

Table 2. Example comment, self-interpretation, and inferences

| Example USER COMMENT |  |
| --- | --- |
| "I like this charger, it has a long battery life and it is small so that it does not take up too much space in my backpack. The only problem is that it leaks battery so if it not in use for a while, it runs out of charge even though I have not used it." | |
| **User self-interpretation baselines** | |
| Do-goals | I aim to charge my phone on the go. I aim to be able to travel with this portable charger and that it does not take up too much space. |
| Motor-goals | I need to make sure my portable charger is fully charged before my travel. I need to keep a cable with it so that I can connect it to my phone. I should be able to charge for 48 hours before the battery runs out. |
| Be-goals | I aspire to stay connected. I need to feel safe in knowing that I have a means of contacting someone. |
| **Example of DESIGNER INFERENCE** | |
| Do-goals (0.789) | I aim to charge my devices (phone, laptop, etc.) at any time and in any place when they are out of battery; I aim to store that ability to charge in a way that does not take up a lot of space; I aim to be able to charge my devices at any time for a long period before I need to recharge the charger. |
| Motor-goals (0.542) | By plugging my devices into the charger, I should be able to recharge their batteries; By placing the charger in my backpack, I should be able to store it and access it conveniently again in the future; By removing the charger from my backpack and plugging it into a socket, I should be able to refill the capacity of the charger so I can use it again for a long period of time. |
| Be-goals (0.571) | I aspire to be secure and not be caught in a position where my devices are out of battery and cannot be used; I aspire to achieve whatever my devices motivate me to achieve: being connected by using my phone; having an impact and being competent by doing my work on my laptop; being stimulated using those electronics at any given time; I aspire to be autonomous and not have to rely on someone else or be tethered to a specific location to be able to use my electronics. |
| **Example of LLM INFERENCE (Standard, GPT-4)** | |
| Do-goals (0.797) | I aim to keep my devices charged while on the go; I aim to conserve space in my backpack. |
| Motor-goals (0.580) | By regularly charging the portable charger, I can ensure it's ready to charge my devices when needed; By placing the charger in a small compartment or pocket in my backpack, I can save space for other items. |
| Be-goals (0.586) | I aspire to be always connected and reachable; I aspire to be organized and efficient in my daily activities. |
| **Example of LLM INFERENCE (ToT, GPT-4)** | |
| Do-goals (0.673) | Keep my electronic devices charged during travel or daily use, particularly when I don't have immediate access to a power outlet. Keep my backpack compact and not overloaded with bulky items. |
| Motor-goals (0.631) | Ensure that the portable charger is fully charged before I pack it into my backpack. Regularly check the charger's battery status to ensure it hasn't drained due to the leakage issue. If the battery is low, recharge it before I can use it again. |
| Be-goals (0.363) | Be self-reliant and prepared, not being caught off-guard by a dead battery in my device. Value efficiency and compactness, minimizing the clutter and weight in my backpack. |

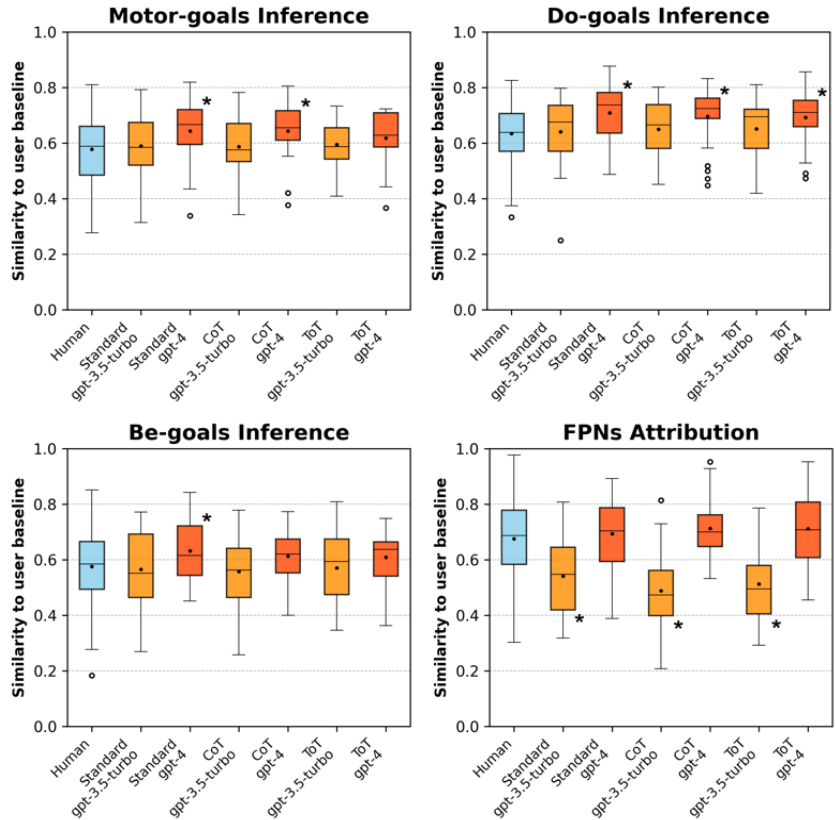

**Figure 4. Inference performance across all groups, boxes marked with an asterisk (*) signify statistical differences from human performance, achieving a significance level of 5%.**

Next, we investigated the relationship between the mental inference performances of both human designers and the LLM with the length of comment that they are asked to analyze. This investigation is based on a simple but reasonable assumption: the more information the user provides, the easier it will be for designers and LLM to take the user's perspective. The comment texts are tokenized and stop-words-excluded using the Natural Language Toolkit (NLTK). The token length after processing more accurately reflects the amount of information carried with the comments. The results are shown in figure 5. For each group, we calculate Spearman's correlation coefficients relative to comment token length to investigate the presence of any linear correlations. The only group where a weak correlation is found is the gpt-3.5-turbo under standard prompting, when attributing FPNs, with a significance level of 7.82%. However, as analyzed previously, this group shows a lack of accuracy in correctly attributing FPNs. Therefore, contrary to expectations, no significant correlation was found between the length of the comment and mental inference performance scores. One possible explanation of this result may be that much of the required information for inference is coming from the inferrers' previous knowledge or experiences [5, 6]. Another explanation could be that even the longest comments we collected did not contain sufficiently richer information to enhance inference performance.

An Investigation into LLM-based Empathic Mental Inference (A PREPRINT)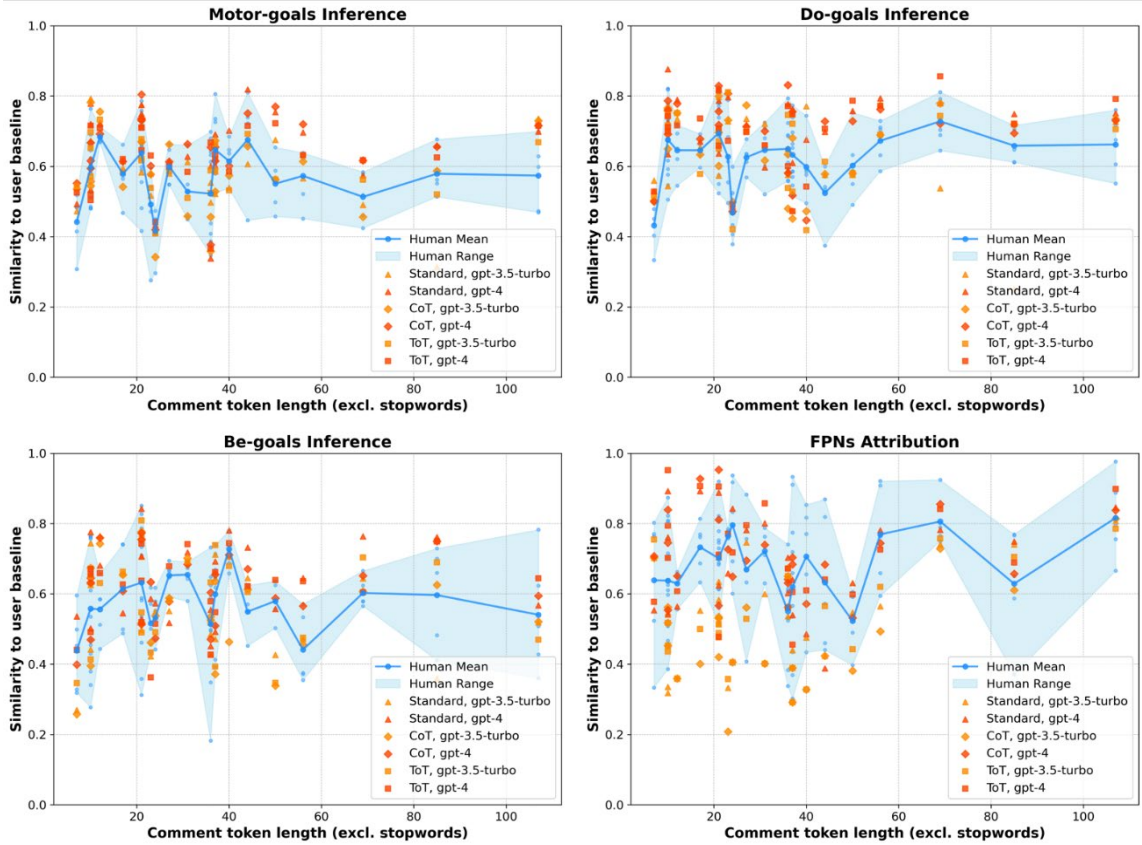

**Figure 5. Inference performance over comment token length**

## 5  Discussion

Overall, the experiment results suggest that the mental inference performance of LLMs can match, and in some cases even surpass, that of human designers. This finding addresses our research question posed in Section 1. The results affirm the potential utility of LLMs in the mental inference component of AE for human-centered design. More immediately, they demonstrate LLMs' capability for analyzing big data on user opinions to discern motivational mental states, going beyond mere explicit retrieval and extraction. Furthermore, this ability suggests that cognitively demanding tasks in human-centered design, especially those requiring empathic capabilities previously thought exclusive to human designers, could potentially be automated by LLMs.

Despite these promising results, several limitations within our study warrant attention and highlight avenues for future research:

*Limited Sample Size and Diversity*: The experiment was conducted with 24 user participants and 20 designer participants, the majority of whom are students or faculty members of Massachusetts Institute of Technology. This concentration potentially restricts the diversity of our dataset, limiting the generalizability of our findings. Future studies should aim to include participants from a broader range of educational and cultural backgrounds to enrich the data and insights derived.

*Constrained Range of User Activities*: The data collection was based on comments pertaining to just three activity categories. This can result in the absence of some aspects of user experience. For example, the FPN for morality is not explicitly involved in all three categories and thus we cannot verify the models' validity to infer this aspect of user motivation.

*Inherent Limitation in Empathic Accuracy*: The empathic accuracy metric used in our study is adapted from methods employed in empathic design research, comparing the similarity between users' self-reported mental states and designers' inferred ones. However, this approach may not fully capture the nuances of empathic understanding. Inferrers might accurately identify mental states that users themselves are not consciously aware of to be able to express them [15], leading to potentially insightful but technically inaccurate inferences according to empathic accuracy. Developing a follow-up evaluation metric that includes user judgement on the inferences could provide a more nuanced assessment of empathic accuracy.

Looking forward, one of the primary objectives of Artificial Empathy (AE) research in human-centered design, as discussed by Zhu and Luo [14], is to develop both large-scale and in-depth empathic understanding to support design research and practice. This is crucial for designing solutions that meet the real needs of a large human population, a goal that is currently challenging to achieve. This study represents an initial step towards achieving that goal by examining the capabilities of AE in understanding the motivation-directed mental contents of users' goals and FPNs. However, we acknowledge that focusing solely on goals and needs may not capture the full complexity of human mental states. Future research should explore a wider variety of mental states (e.g., emotional appraisals) and incorporate multimodal inference to further advance the field of AE.

## 6   Conclusion

In the presented paper, we proposed an LLM-based method to perform empathic mental inference tasks based on user comments and conducted human subject study to develop an empathic accuracy metric for measuring its performance. The results suggest the LLM-based approach can achieve promising performance comparable to human designers doing the same tasks.